\documentclass[twocolumn,english,aps,prb,showpacs,byrevtex,amsmath,amssymb,superscriptaddress]{revtex4-1}

\usepackage{url} 
\usepackage{graphicx}
\usepackage{epstopdf}
\usepackage{dcolumn}
\usepackage{bm}
\usepackage{gensymb}
\usepackage{upgreek}
\usepackage{hyperref}
\usepackage{graphicx}
\usepackage{epstopdf}
\usepackage{dcolumn}
\usepackage{bm}
\usepackage{gensymb}
\usepackage{upgreek}
\usepackage{booktabs}
\usepackage{hyperref}

\usepackage{amssymb,mathtools}
\usepackage{color}
\usepackage{float}

\usepackage{tikz,xcolor,hyperref}
\definecolor{lime}{HTML}{A6CE39}
\DeclareRobustCommand{\orcidicon}{%
	\begin{tikzpicture}
	\draw[lime, fill=lime] (0,0)
	circle [radius=0.16]
	node[white] {{\fontfamily{qag}\selectfont \tiny ID}};
	\draw[white, fill=white] (-0.0625,0.095)
	circle [radius=0.007];
	\end{tikzpicture}
	\hspace{-2mm}
}

\foreach \x in {A, ..., Z}{%
	\expandafter\xdef\csname orcid\x\endcsname{\noexpand\href{https://orcid.org/\csname orcidauthor\x\endcsname}{\noexpand\orcidicon}}
}

\begin{document}

\title{Electronic and optical properties of InAs/InAs$_{0.625}$Sb$_{0.375}$ superlattices\\ and their application for far-infrared detectors}

\author{Ghulam Hussain\orcidE}
\email{ghussain@magtop.ifpan.edu.pl}
\affiliation{International Research Centre MagTop, Institute of Physics, Polish Academy of Sciences, Aleja Lotnik\'ow 32/46, PL-02668 Warsaw, Poland}

\author{Giuseppe Cuono\orcidD}
\email{gcuono@magtop.ifpan.edu.pl}
\affiliation{International Research Centre MagTop, Institute of Physics, Polish Academy of Sciences, Aleja Lotnik\'ow 32/46, PL-02668 Warsaw, Poland}

\author{Rajibul Islam\orcidC}
\affiliation{International Research Centre MagTop, Institute of Physics, Polish Academy of Sciences, Aleja Lotnik\'ow 32/46, PL-02668 Warsaw, Poland}
\author{Artur Trajnerowicz}
\affiliation{VIGO System S.A. 129/133 Poznanska Str., 05-850 Ozarow Mazowiecki, Poland}
\author{Jaros{\l}aw Jure\'nczyk}
\affiliation{VIGO System S.A. 129/133 Poznanska Str., 05-850 Ozarow Mazowiecki, Poland}
\author{Carmine Autieri\orcidA}
\email{autieri@magtop.ifpan.edu.pl}
\affiliation{International Research Centre MagTop, Institute of Physics, Polish Academy of Sciences, Aleja Lotnik\'ow 32/46, PL-02668 Warsaw, Poland}
\affiliation{Consiglio Nazionale delle Ricerche CNR-SPIN, UOS Salerno, I-84084 Fisciano (Salerno),
    Italy}
\author{Tomasz Dietl\orcidB}
\affiliation{International Research Centre MagTop, Institute of Physics, Polish Academy of Sciences, Aleja Lotnik\'ow 32/46, PL-02668 Warsaw, Poland}
\affiliation{WPI-Advanced Institute for Materials Research, Tohoku University, Sendai 980-8577, Japan}

\begin{abstract}
We calculate the electronic and optical properties of InAs/InAs$_{0.625}$Sb$_{0.375}$ superlattices within relativistic density functional theory. To have a good description of the electronic and optical properties, the modified Becke-Johnson exchange-correlation functional is pondered to correctly approximate the band gap. First, we analyze  electronic and optical characteristics of bulk InAs and InSb, and then we investigate the InAs/InAs$_{0.625}$Sb$_{0.375}$ superlattice.
The optical gaps deduced from the imaginary part of the dielectric function are associated with the characteristic interband transitions. 
We investigate the electronic and optical properties of the InAs/InAs$_{0.625}$Sb$_{0.375}$ superlattice with three lattice constants of the bulk InAs, GaSb and AlSb, respectively. It is observed that the electronic and optical properties strongly depend on the lattice constant. Our results support the presence of two heavy-hole bands with increasing in-plane effective mass as we go far from the Fermi level. We notice a considerable decrease in the energy gaps and the effective masses of the heavy-holes in the k$_x$-k$_y$ plane compared to the bulk phases of the parent compounds. We demonstrate that the electrons are s-orbitals delocalized in the entire superlattice, while the holes have mainly 5p-Sb character localized in the In(As,Sb) side of the superlattice. In the superlattice, the low-frequency absorption spectra greatly increase when the electric field is polarized orthogonal to the growth axis allowing the applicability of III-V compounds for the long-wavelength infrared detectors. 

\end{abstract}

\date{\today}
\maketitle

\section{Introduction}

Over the last few years,  InAs/In(As,Sb) Ga- and Hg-free type-II superlattices (T2SLs) have attracted attention as a promising candidate for infrared detector applications due to their narrow gaps and carrier lifetime longer than in other systems \cite{Rogalski17InAs,Rogalski20,Ting20a,Ting20b,Martyniuk20,Ting19,Wu19,Steenbergen11,Klipstein14}.
Furthermore, In(As,Sb) quantum wells and nanowires have been proposed  as a platform for topological superconductivity  \cite{moehle2021insbas} while the InAs/GaSb heterostructures host topological states due to their type III band bending\cite{PhysRevLett.112.176403,Winkler16Topological}. Typically, the band structure of these compounds is studied by the envelope function $\bm{k}\cdot\bm{p}$
formalism or within tight-binding approximation\cite{Rogalski20}. These empirical methods are successful in describing the conduction band but often fail in the case of the heavy-hole valence band, whose curvature and anisotropy are determined
by remote bands not considered usually within effective hamiltonian methods. In this paper, we determine, by a first-principle method, band structures and optical properties of bulk In(As,Sb) and InAs/InAs$_{0.625}$Sb$_{0.375}$ SLs with lattice constants of InAs, GaSb and AlSb, respectively.
Our results show that the absorption coefficient of SLs is larger than that of bulk materials, reconfirming the superiority of SLs for detector applications. We demonstrate also that the effective mass of heavy-holes in the k$_x$-k$_y$ plane is much reduced in SLs indicating that
III-V SLs might replace IV-VI compounds as efficient Pb-free infrared emitters.

\noindent The InAs and InSb bulk compounds belong to the family of the III-V zinc-blende semiconductors.
For the zinc-blende systems and their alloys, it is possible to grow heterostructures and superlattices, in which the alloy content, thickness and interface chemistry can serve to tune the band gap and bandwidths\cite{Mikhailova_2004} of the two sides of the interface to have type II or type III SL.
In a first approximation, in the InAs/In(As,Sb) SLs due to the type II band alignment the electrons are confined
within the InAs layer and the holes in the In(As,Sb) layers, and thus electrons and holes are spatially
separated. Therefore, by adjustment of the InAs and/or In(As,Sb) thickness as well as the Sb concentration, it is feasible to tune the band gap within a wide range of the infrared region.
Moreover, the thickness and the Sb concentration have to be tuned to keep the average lattice constant of SL as close as possible to the lattice constant of the substrate.
Owing to these unique properties, T2SLs InAs/In(As,Sb) have been chosen as materials for applications in the far-infrared radiation. 
In this paper, we will give more details about the location and dispersion of electrons and holes in T2SLs InAs/In(As,Sb).

\begin{figure*}[]
	\centering
	\includegraphics[scale=0.6, angle=0]{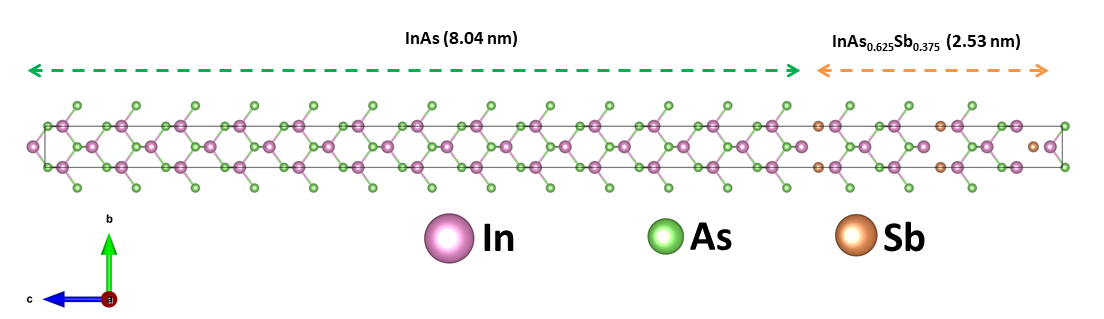}
	\caption{Optimized crystal structure of the InAs/InAs$_{0.625}$Sb$_{0.375}$ SL grown along the c-axis, (001) axis of the bulk unit cell. The orchid, green and sandy brown spheres represent In, As and Sb atoms, respectively. The dashed green arrow represents the extension of the InAs side of the supercell, while the dashed orange arrow represents the extension of the InAs$_{0.625}$Sb$_{0.375}$ side of the superlattice. The solid black line represents the unit cell of the superlattice.}
	\label{Figure1}
\end{figure*}

In the next Section, we describe the computational details, the third Section is devoted to the electronic and optical properties of the InAs and InSb bulk while in the fourth Section the InAs/InAs$_{0.625}$Sb$_{0.375}$ SL properties are investigated. Finally, the last Section is dedicated to the conclusions.

\section{Computational details}

We have performed DFT calculations by using the VASP package \cite{Kresse93,Kresse96a,Kresse96b}.
The core and the valence electrons were treated within the projector augmented wave (PAW) method\cite{Kresse99} with a cutoff of 300 eV for the plane wave basis. Given the importance of  spin-orbit coupling (SOC) for the band gap in zinc blende and given the size of the SOC of those elements, all calculations were performed with the relativistic effects taken into account. For compounds with a zinc blende structure, DFT within a standard local spin density approximation does not produce good electronic properties, e.g., band gaps, spin-orbit splittings, and effective masses.
The resulting band structure has a wrong band ordering at the $\Gamma$ point; these compounds turn out to be zero-gap topological semimetals (like HgTe) rather than narrow-gap semiconductors.
Few approaches have made it possible to solve this problem, among these, there is the hybrid exchange-correlation functional\cite{kim2009accurate,Sliwa22} and an all-electron screened exchange approach within the full potential linearized augmented plane-wave method \cite{Geller01}.
These approaches reproduce the experimental band gaps of these compounds, however, they are computationally expensive to study heterostructures.
Therefore, in order to study heterostructures and superlattices, we have employed the modified Becke-Johnson (MBJ) exchange-correlation functional together with generalized gradient approximation (GGA) for the exchange-correlation potential\cite{Tran09}, which is a semilocal potential that improves the description of the band gaps especially for narrow gap semiconductors\cite{Camargo12,Islam19,Goyal17} including zinc blende semicondutors\cite{Autieri21,sliwa2021superexchange,Sliwa22,Islam21a}.
We can tune the band gap by varying the parameter $c^{\text{MBJ}}$ for the different atoms, as reported in the next section. The value of $c^{\text{MBJ}}$ is the only adjustable parameter that is fixed fitting the band gap.
In the literature, the MBJ approach was successfully used for other zinc-blende systems\cite{Silva19}.
Despite the MBJ functional having problems with doped systems due to its non-locality\cite{Rauch20Local}, we have found that our results for the bulk system
are quite remarkable. Therefore, we conclude that the MBJ provides reliable results if the properties of the doping do not differ too much from the properties of the host as in the case InAs doped with Sb. 
The GGA+U does not solve the gap problem for the zinc blende unless we would use exotic negative Coulomb repulsions\cite{Yu2020}.  \\

We have performed the bulk calculations using 10$\times$10$\times$10 $k$-points centred in $\Gamma$.
The values of the lattice parameters at $T=0$\,K are: $a_{\text{InAs}} = 6.05008$\,{\AA} and $a_{\text{InSb}} = 6.46896$\,{\AA} \cite{Vurgaftman01}.
Due to the large value of the out-of-plane SL period, the k$_z$ range in the SL first Brillouin zone is rather small. 
We have carried out the InAs/InAs$_{0.625}$Sb$_{0.375}$ SL calculations using 10$\times$10$\times$1 $k$-points centred at $\Gamma$ .
Before calculating the electronic and optical properties of the SL, we have performed the structural relaxation of the internal degrees of freedom within GGA without SOC. The reliability of the structural relaxation without SOC was proved in the literature\cite{Kriegner11} and tested by us for simple cases.  
To inspect how the energy gaps and optical properties of InAs/InAs$_{0.625}$Sb$_{0.375}$ SL change as a function of the lattice constant, we used room temperature lattice constants of InAs, GaSb and AlSb, respectively. The values of the lattice parameters at $T=300$\,K are: $a_{InAs}$=6.0583 {\AA}, $a_{GaSb}=6.0959$\,{\AA} and $a_{AlSb}=6.1355$\,{\AA} \cite{Vurgaftman01}.
In other words, we considered the SL is grown on InAs, GaSb and AlSb substrates, respectively.
The SLs are treated as fully strained and we have a perfect interface without intermixing.
Our (001) SL is composed of 8.04\,nm of InAs (26 unit cells) and 2.53 nm of InAs$_{5/8}$Sb$_{3/8}$ (8 unit cells), as shown in Fig.~\ref{Figure1}.
Thus, the supercell size along the growth direction is 10.57 nm.
We use the lattice constants and the optimal doping concentration from the experimental system.
We construct a supercell with a single atom species in the subsequent layers, so that the InAs$_{5/8}$Sb$_{3/8}$ consists of eight unit cells, with three Sb and five As layers, respectively.
We performed three different calculations for the SL using InAs, GaSb and AlSb in-plane lattice constants at room temperature.
This allows us to investigate the properties of SL close to the substrate and the fully relaxed one far from the substrate.

The optical properties are obtained by using the frequency-dependent dielectric function $\epsilon (\omega)$ written as:
\begin{equation*}
\epsilon (\omega)
 = \epsilon_1 (\omega) + i\epsilon_2 (\omega)
\end{equation*}
where $\omega$ is the photon angular frequency, while $\epsilon_1 (\omega)$ and  $\epsilon_2 (\omega)$ represent the real and imaginary parts of the complex dielectric function, respectively. 
The imaginary part can be calculated in random phase approximation neglecting local field effects via the output of the DFT calculations using the following formula:
\begin{gather} \nonumber
\epsilon_2 (\omega)=\frac{\nu e^2}{2\pi h m^2 \omega^2}\int d^3k\sum_{mn'} f_{n,\mathbf{k}} (1-f_{n',\mathbf{k}})\\
|<n,\mathbf{k}|\mathbf{p}|n',\mathbf{k}>|^2\times\delta(E_{n,\mathbf{k}} - E_{n',\mathbf{k}} - \hbar\omega)
\end{gather}
where f is the Fermi-Dirac distribution, $\mathbf{p}$ is the momentum operator, $E_{n,\mathbf{k}}$ is the energy spectrum as a function of the number of bands $n$ and wave vector $\mathbf{k}$.
The real part of the optical conductivity can be calculate using the Kramers-Kronig transformation: 
\begin{equation}
\epsilon_1 (\omega)= 1 + \frac{2}{\pi}P\int_{0}^{\infty} \frac{\epsilon_2 (\omega')\omega'}{\omega'^2 - \omega^2} d\omega'
\end{equation}
This approach is known to overestimate the dielectric constant\cite{PhysRevB.73.045112}, however the large supercell and the inclusion of the SOC does not allow us to go beyond.
The absorption coefficient $\alpha(\omega)$ can be derived from  the $\epsilon_1 (\omega)$ and $\epsilon_2 (\omega)$ as follows:
\begin{equation*}
 \alpha = \frac{\substack{\sqrt{2}\omega}}{\substack{c}} \left( \sqrt{\epsilon_1^2 + \epsilon_2^2} - \epsilon_1 \right)^{1/2}
\end{equation*}
where $c$ is the speed of light.
We employed the independent-particle approximation, setting the number of points in the energy range of the density of states NEDOS=30000 for the bulk and NEDOS=40000 for the SL. This approach allows us to perform accurate optical calculations for the superlattice within a reasonable computational cost.
The optical properties were already calculated successfully using MBJ for other semiconductors like MnS, GaN, Si, MnO, GaAs etc.~\cite{tran2009accurate}.
\\

 \begin{figure}[]
	\centering
	\includegraphics[scale=0.22, angle=0]{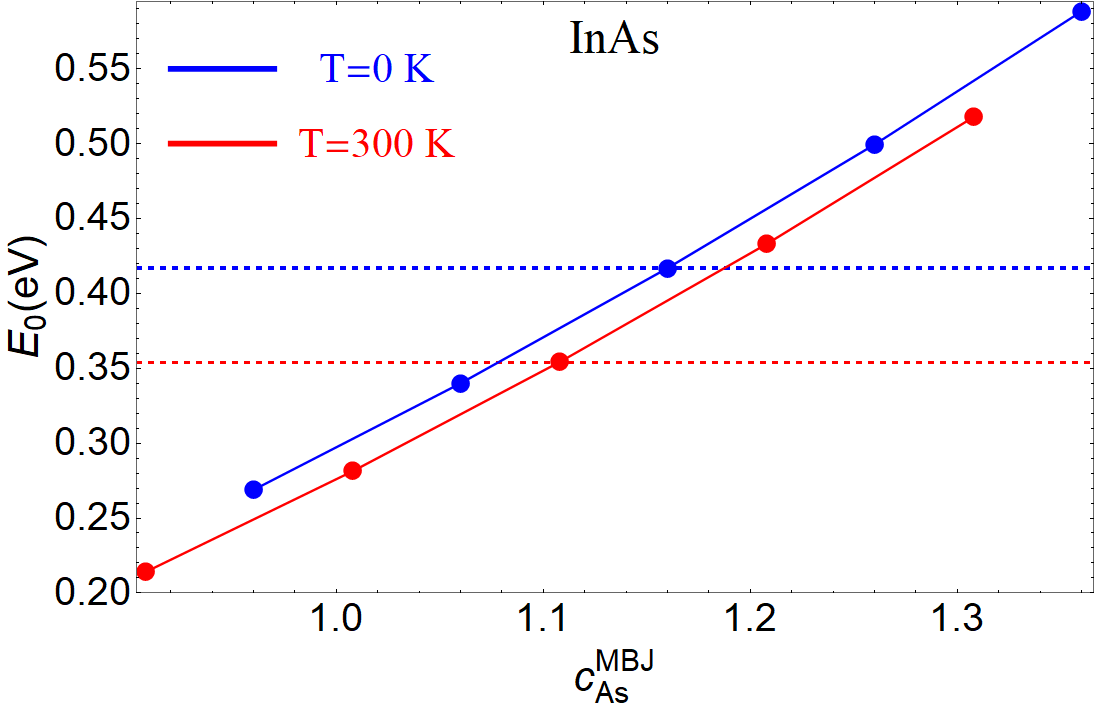}
	\caption{Band gap at $\Gamma$ point for InAs as a function of c$^{MBJ}_{As}$, and with c$^{MBJ}_{In} =1.180$. The blue line indicates the band gap at $T=0$\,K, while the red line the band gap at $T=300$\,K. The experimental band gap is represented by the horizontal blue dashed line in the case of $T=0$\,K, while by the horizontal red dashed line in the case of $T=300$\,K.}
	\label{Figure2}
\end{figure}

\begin{figure}[]
	\centering
	\includegraphics[scale=0.22, angle=0]{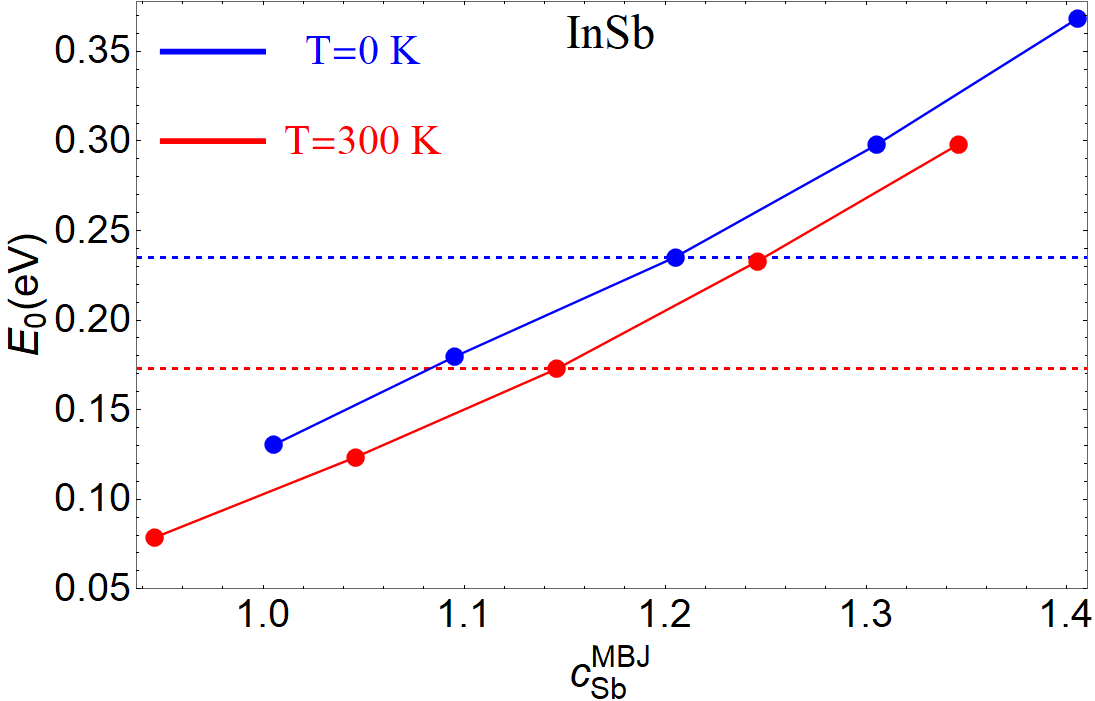}
	\caption{Band gap at $\Gamma$ point for InSb as a function of c$^{MBJ}_{Sb}$, and with c$^{MBJ}_{In}$=1.180. The blue line indicates the band gap at T=0 K, while the red line the band gap at T=300 K. The experimental band gap is represented by the horizontal blue dashed line in the case of T=0 K, while by the horizontal red dashed line in the case of T=300 K.}
	\label{Figure3}
\end{figure}

 \begin{figure*}[]
	\centering
	\includegraphics[scale=0.6,angle=0] {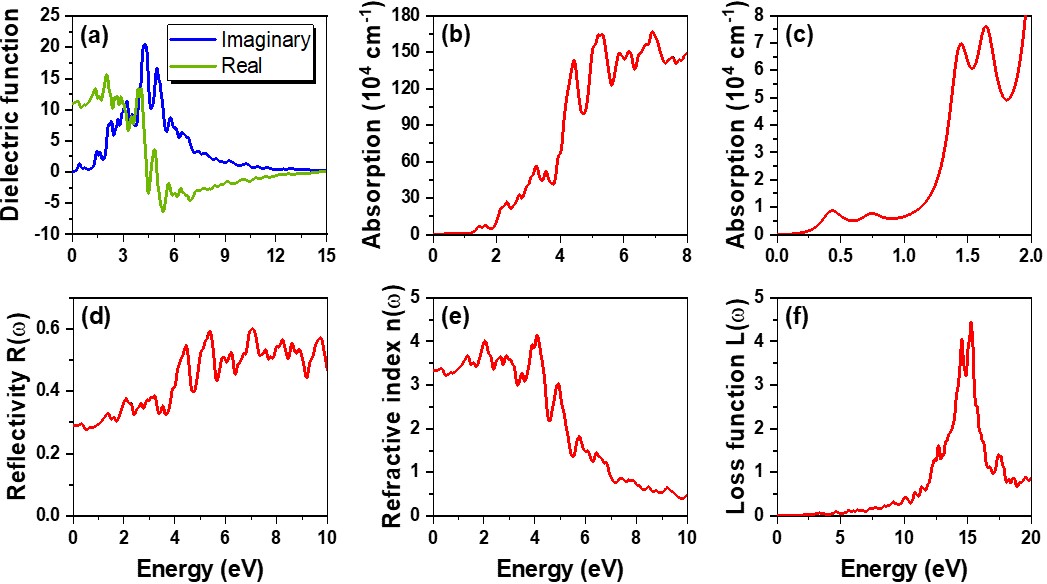}
	\caption{Frequency-dependent optical properties of InAs bulk with the low temperature lattice constant. (a) Real (green line) and imaginary part (blue line) of the dielectric function. (b)  Absorption spectra in the frequency-range between 0 and 8 eV. (c) Absorption spectra in the IR region. (d) Reflectivity. (e) Refractive index. (f)  Energy loss function.}
	\label{Figure4}
\end{figure*}

\section{Bulk properties}

\begin{table*}[th]
\centering
\caption{Effective masses of bulk InAs calculate within 
MBJ (this work) compared with other theoretical approaches\cite{kim2009accurate} and experimental results\cite{shur1996handbook, chen1995semiconductor} at $\Gamma$ along different direction of the \textbf{k}-space. Increasing the energy, we name the bands as split-off, light-hole, heavy-hole and electron band as used in the literature.\cite{kim2009accurate}}
\setlength{\tabcolsep}{8 pt}
\renewcommand{\arraystretch}{1.5}
\scalebox{1.2}{
\label{Table 1}
\begin{tabular}{cccccc}
\hline
Approach & Direction & m$_{split-off}$ & m$_{light-hole}$ & m$_{heavy-hole}$ & m$_{electron}$ \\
\hline

HSE06  & [100] & 0.112 & 0.033 & 0.343 & 0.027\\

 & [111] & 0.111 & 0.031 & 0.836 & 0.027\\

 & [110] & 0.112 & 0.032 & 0.623 & 0.027\\
\hline
MBJ & [100] & 0.189 & 0.064 & 0.155 & 0.021\\

 & [111] & 0.119 & 0.046 & 0.609 & 0.014\\

 & [110] & 0.062 & 0.037 & 0.409 & 0.031\\
 
 \hline
EXPERIMENT & [100] & 0.140 & 0.027 & 0.333 & 0.023\\

 & [111] & 0.140 & 0.037 & 0.625 & 0.026\\

 & [110] & 0.140 & 0.026 & 0.410 & 0.026\\

\hline
\end{tabular}}
\end{table*}

\begin{table}[th]
\centering
\caption{Static dielectric constants  $(\varepsilon)$ and refractive indices (n$_{o}$) for the InAs bulk in the present work and in the literature.}
\setlength{\tabcolsep}{15 pt}
\renewcommand{\arraystretch}{1.5}
\scalebox{1.0}{
\label{Table 2}
\begin{tabular}{ccc}
\hline
Approach & $\varepsilon$ & (n$_{o}$) \\
\hline
Present work & 11.07  & 3.315 \\
GGA-EV theory \cite{sohrabi2017structural} & 10.82  & 3.289 \\
Experiment \cite{philipp1963optical} & 11.67  & - \\
\hline
\end{tabular}}
\end{table}

 \subsection{Electronic properties of the InAs and InSb bulk}

For the bulk, we used the MBJ exchange-correlation functional wherein the parameter $c^{\text{MBJ}}$ was adjusted to reproduce the experimental band gaps at the $\Gamma$ point, which  are 0.417\,eV for InAs and 0.235\,eV for InSb, respectively at temperature $T=0$,\cite{Vurgaftman01} and  decrease to 0.354\,eV for InAs and 0.173\,eV for InSb at 300 \,K.\cite{Madelung02}
From our calculations, we noticed that the gap is rather sensitive to the value of the lattice parameter.
We can choose one value of $c^{\text{MBJ}}$ for each element. We fix $c^{\text{MBJ}}_{\text{In}}=1.180$, so we can adjust the magnitudes of the single parameters $c^{\text{MBJ}}_{\text{As}}$ and $c^{\text{MBJ}}_{\text{Sb}}$ in order to reproduce the experimental values of the band gap for InAs and InSb, respectively. We repeat the procedure at 0 and 300\,K.
In Figs.~\ref{Figure2} and \ref{Figure3} we show the computed band gap at the $\Gamma$ point for InAs and InSb as a function of $c^{\text{MBJ}}_{\text{As}}$ and  $c^{\text{MBJ}}_{\text{Sb}}$ for experimental lattice parameters at $T=0$ and 300\,K, respectively.
As seen, the experimental band gaps are reproduced assuming $c^{\text{MBJ}}_{\text{As}}=1.160$ for InAs and  $c^{\text{MBJ}}_{\text{Sb}} =1.205$ for InSb at 0\,K while we need $c^{\text{MBJ}}_{As}=1.108$ for InAs and $c^{\text{MBJ}}_{\text{Sb}}=1.146$ for InSb at 300\,K.
In this way, we have obtained information on the behaviour and the  possible range of $c^{\text{MBJ}}$ for the bulk systems.

The band bending of the InAs/InSb returns a type III SL, while in virtual crystal approximation the band bending of the InAs/InAs$_{0.625}$Sb$_{0.375}$ returns a type II SL in agreement with experimental data. This highlights how our computational setup is reliable in the description of the electronic properties of this material class.

Using the notation of reference \onlinecite{kim2009accurate}, Table \ref{Table 1} summarizes various theoretical and experimental approaches to approximate the effective masses of the energy bands such as m$_{split-off}$, m$_{light-hole}$, m$_{heavy-hole}$ and conduction electron m$_{electron}$ along the directions [100], [110] and [111]. The effective masses calculated for heavy-hole along the [111] and [110] directions strongly match experimental results. Similarly, the effective masses of conduction electrons in [100] and [111] directions coincide with those of experiments. Respect to the Heyd–Scuseria–Ernzerhof (HSE06) exchange-correlation, the light-holes are heavier and the heavy-holes are lighter in MBJ.

\begin{figure*}[t]
	\centering
	\includegraphics[scale=0.6, angle=0] {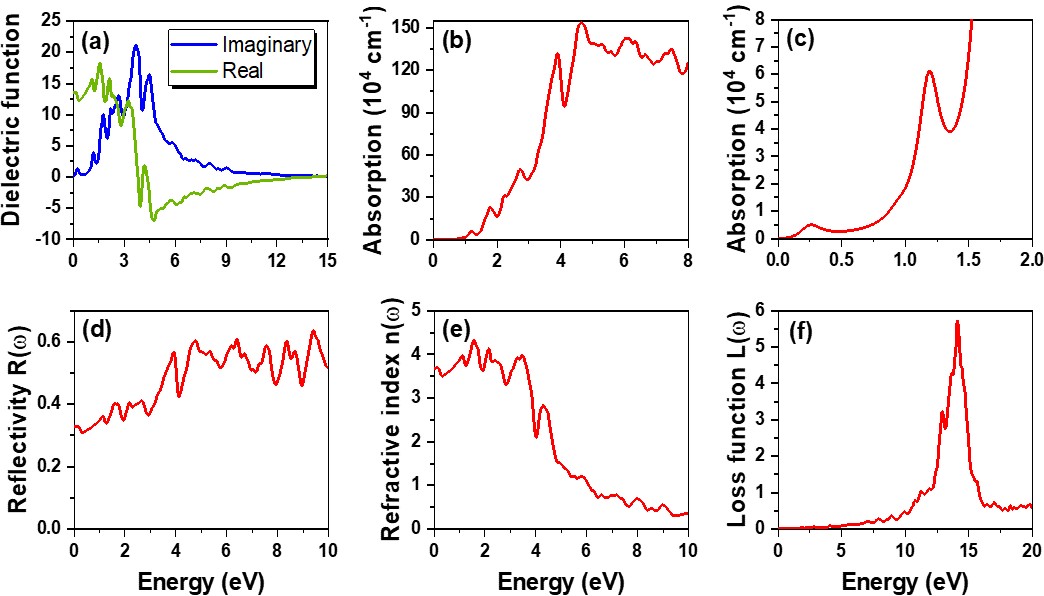}
	\caption{Frequency-dependent optical properties of InSb bulk with the low temperature lattice constant. (a) Real (green line) and imaginary part (blue line) of the dielectric function. (b)  Absorption spectra in the frequency-range between 0 and 8 eV. (c) Absorption spectra in the IR region. (d) Reflectivity. (e) Refractive index. (f) Energy loss function.}
	\label{Figure5}
\end{figure*}

\subsection{Optical properties of the InAs and InSb bulk}
The optical properties of a particular semiconductor are of great interest due to their possible applications in optoelectronics and photon sensing. After establishing the values of $c^{\text{MBJ}}$, we calculate the optical properties of InAs and InSb at T=0.  In Figs.~\ref{Figure4} and \ref{Figure5} we show  optical characteristics of InAs and InSb in the bulk form, respectively. Due to the cubic symmetry, the optical properties of InAs an InSb bulk are isotropic and therefore independent on whether the electric field is polarized along the x-, y- and z-axis.
Figs.~\ref{Figure4}(a) and \ref{Figure5}(a) illustrate the real and imaginary parts of the dielectric function for InAs and InSb, which are used to extract other optical responses for the system. The values of the real part of the dielectric function on the vertical y-axis of the graph are known as static values of the real part; the value of static dielectric constant for InAs is 11.07, while that of InSb is 13.42, which is in agreement with the values reported in theoretical \cite{sohrabi2017structural, ziane2014first} and experimental works \cite{philipp1963optical}. Table \ref{Table 2} gives a comparison between the present work and the literature. The imaginary and real parts of the dielectric function\cite{PhysRevB.27.985} as well as the refractive index are in very good agreement with the literature\cite{website_refractive} for both InAs and InSb.\\ 

If we look at the absorption spectra of Figs.~\ref{Figure4}(b) and \ref{Figure5}(b), we can see that the first absorption peaks appear at 0.417\,eV for InAs and 0.235\,eV for InSb. These peaks correspond to the energy-gap between the highest occupied and the lowest unoccupied energy bands, characterizing specific electronic transitions. In Figs.~\ref{Figure4}(c) and \ref{Figure5}(c) we zoom in the 0-2 eV energy range, to quantify and describe the importance of InAs and InSb semiconductors in the infrared (IR) region.
Previous calculations on the same material class were reported using all electron method with $c^{\text{MBJ}}$ calculated self-consistently\cite{NAMJOO2022,Namjoo2015} and tight-binding method\cite{Smith1987Proposal}, despite the different methodology our results are in quantitative agreement with the theoretical literature.
If we compare our absorption spectra with experimental literature, we find that our theoretical absorption coefficients overestimate the experimental values as expected neglecting the local field effect. 

\begin{table*}[th]
\centering
\caption{Effective masses of InAs/InAs$_{0.625}$Sb$_{0.375}$ at $\Gamma$ point along in-plane directions of the \textbf{k}-space for the lattice constants of InAs, GaSb and AlSb and their comparison with the literature.\cite{Manyk19vigo}}
\setlength{\tabcolsep}{5 pt}
\renewcommand{\arraystretch}{1.5}
\scalebox{1.0}{
\label{Table 3}
\begin{tabular}{cccc|cc|c}
\hline
Lattice constant ({\AA})& Band gap & Direction & m$_{light-hole}$ & m$_{heavy-hole2}$ & m$_{heavy-hole1}$ & m$_{electron}$ \\
\hline
InAs (6.0583) & 0.116 eV & [100] &  & 0.196 & 0.175 & 0.022\\
&   & [110]  & & 0.062 & 0.056 & 0.023\\
\hline
GaSb (6.0959) & 0.087 eV & [100] &  & 0.147 & 0.126 & 0.016\\
&   & [110] & & 0.052 & 0.041 & 0.018\\
\hline
AlSb (6.1355) & 0.053 eV & [100] &  & 0.112 & 0.099 & 0.016\\
&   & [110]  & & 0.046 & 0.044 & 0.024\\
\hline
k${\cdot}$p model for 14.5 nm SL\cite{Manyk19vigo} & 0.1 eV & k$_x$-k$_y$ plane  & 0.096 & \multicolumn{2}{c|}{0.040} & 0.019\\
\hline
InAs/GaSb with 17 layers\cite{RevModPhys.62.173} & 0.1 eV & k$_x$-k$_y$ plane  & 0.051-0.061 & \multicolumn{2}{c|}{0.33-0.37} & 0.023-0.031\\
\hline
\end{tabular}}
\end{table*}

\begin{table*}[th]
\centering
\caption{Effective masses of InAs/InAs$_{0.625}$Sb$_{0.375}$ at $\Gamma$ point along k$_z$ for the lattice constants of InAs, GaSb and AlSb and their comparison with the literature.\cite{Manyk19vigo}}
\setlength{\tabcolsep}{5 pt}
\renewcommand{\arraystretch}{1.5}
\scalebox{1.0}{
\label{Table 4}
\begin{tabular}{cccc|cc|c}
\hline
Lattice constant ({\AA})& Band gap & Direction & m$_{light-hole}$ & m$_{heavy-hole1}$ & m$_{heavy-hole2}$ & m$_{electron}$ \\
\hline
GaSb (6.0959) & 0.087 eV & [001] & & 54.9 & 32.5 & 0.256\\
k${\cdot}$p model for 14.5 nm SL\cite{Manyk19vigo} & 0.1 eV & [001]  & 0.104 & \multicolumn{2}{c|}{31.02} & 0.023\\
\hline
\end{tabular}}
\end{table*}

Besides, the surface behavior of these semiconductors is studied via reflectivity, which is defined as the ratio of the reflected and incident power. Fig.s ~\ref{Figure4}(d) and \ref{Figure5}(d) display the reflectivity spectra for InAs and InSb, which indicate that these compounds are less reflective in the IR region compared to other regions. Nevertheless, the refractive index shows a substantial value in the IR region, as we can see from the Figs. \ref{Figure4}(e) and \ref{Figure5}(e). The low reflectivity and higher refractive index in the IR region make these materials very useful for IR detectors. The static refractive
indices $n_0$ are 3.315 and 3.684 for InAs and InSb, respectively. Further, the energy loss function (ELF) is calculated in Figs. \ref{Figure4}(f) and \ref{Figure5}(f), to measure the loss of energy taking place in the systems. Almost no energy loss can be seen for the photons in the IR region. However, as the energy increases beyond 5 eV, energy loss starts to increase and becomes maximum around ~15 eV. The peaks in ELF spectra correspond to the plasma resonance and hence the associated frequency is the plasma frequency.

\begin{figure*}[t]
	\centering
	\includegraphics[scale=0.55, angle=0] {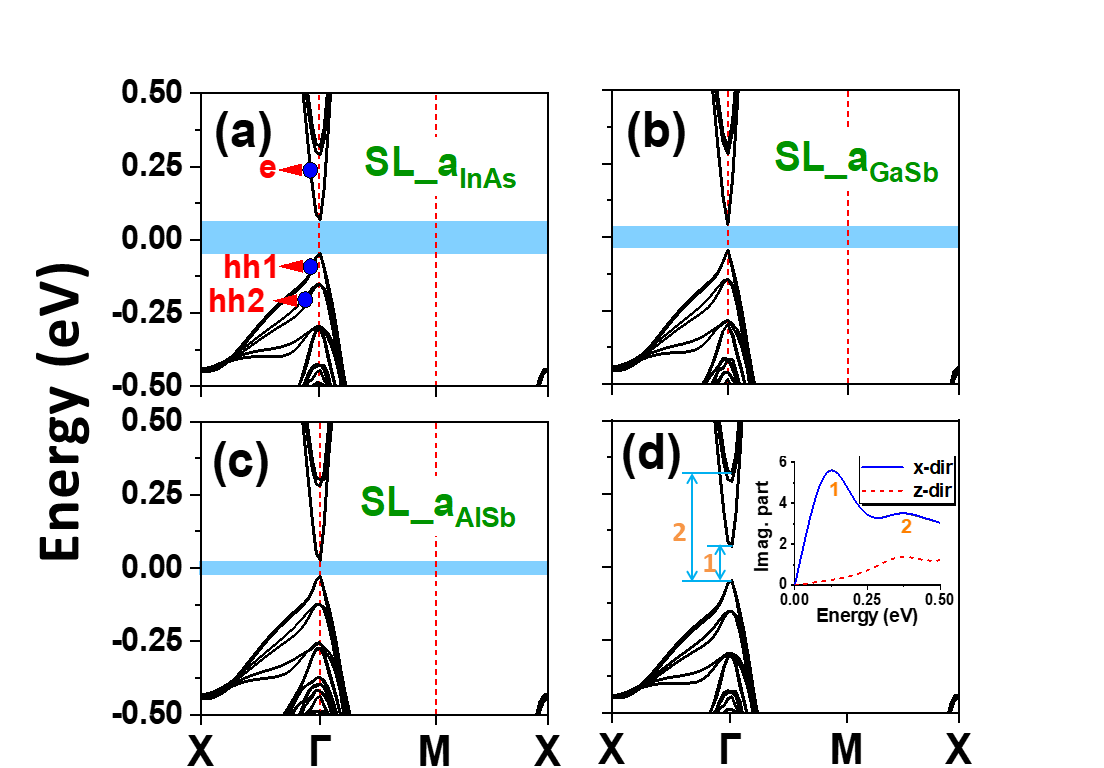}
	\caption{Band structures of the SL for three different lattice constants at 300 K i.e. (a) $a_{InAs}$, (b) $a_{GaSb}$ and (c) $a_{AlSb}$ . (d) Numbers 1 and 2 indicate the characteristic interband transitions. The inset shows the peaks associated with the number 1 and 2 in the imaginary part of the dielectric function for the x- and z-polarization of the electric field. The Fermi level is set to zero.}
	\label{Figure6}
\end{figure*}

\begin{figure}[ht]
	\centering
	\includegraphics[scale=0.37, angle=0] {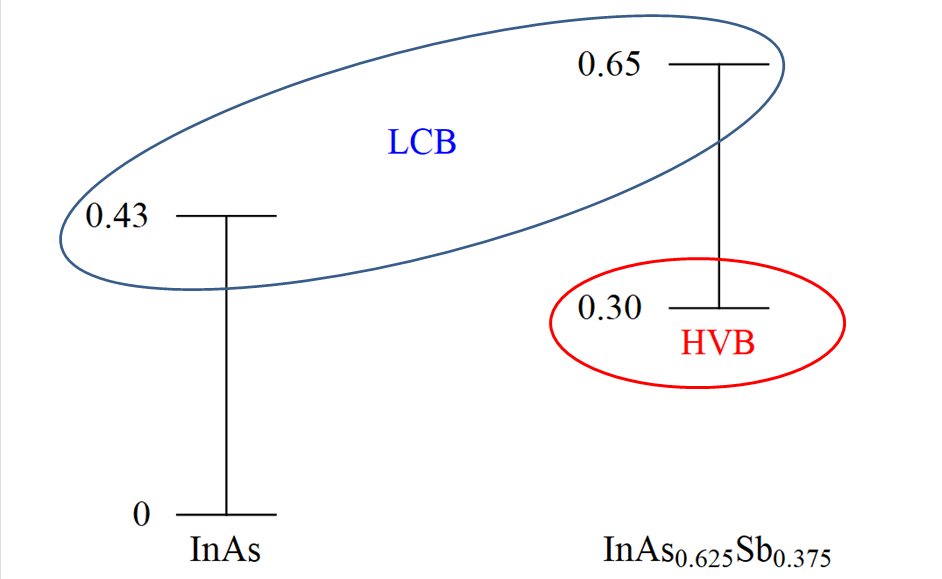}
	\caption{Type-II band banding using the gap of bulk InAs and InAs$_{0.625}$Sb$_{0.375}$ calculated for c$^{\text{MBJ}}$=1.18. From the bulk data, we estimate the LCB and HVB of the supercell. The blue and red circles represent the positions of LCB and HVB in the real space, while the HVB is mainly located in the In(As,Sb) side, the electrons are delocalized along the superlattice. The zero is fixed at the top of the valence band of InAs. The numerical values are in eV.}
	\label{Figure6bis}
\end{figure}

\section{Study of the I\lowercase{n}A\lowercase{s}/I\lowercase{n}A\lowercase{s}$_{0.625}$S\lowercase{b}$_{0.375}$ superlattice}

In this Section, we study the superlattice of InAs/InAs$_{0.625}$Sb$_{0.375}$ since it is employed in the devices of IR-detectors.
Here, we use as in-plane lattice constant of the superlattice the values of the InAs, GaSb and AlSb at $T=300$\,K, respectively\cite{Vurgaftman01}. Using the values of the $c^{\text{MBJ}}$ obtained from the bulk, the superlattice shows a metallic phase, then, we assume that $c^{\text{MBJ}}$ is not a transferable quantity. Therefore, we slightly increase its value and use a single value of $c^{\text{MBJ}}=1.22$ for all atoms. This value gives a reasonable agreement with experimental results of 0.1 eV\cite{Manyk19vigo}. Then, we calculated the band structures and optical properties of the SL for the three different lattice constants. The following two subsections present the electronic and optical attributes with these three lattice constants.

\begin{figure*}[]
	\centering
	\includegraphics[scale=0.6,angle=0] {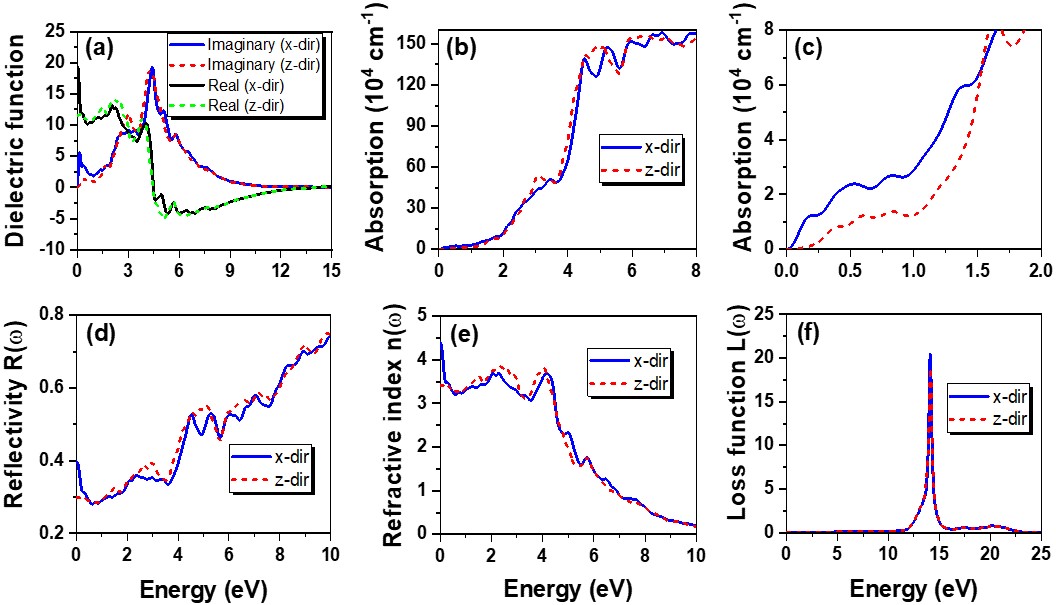}
	\caption{Frequency-dependent optical properties of InAs/InAs$_{0.625}$Sb$_{0.375}$ superlattice with lattice constant at T=300 K that is $a_{InAs}$=6.0583 {\AA}. 
    (a) Real and imaginary part of the dielectric function. (b)  Absorption spectra in the frequency-range between 0 and 8 eV. (c) Absorption spectra in the IR region. (d) Reflectivity. (e) Refractive index. (f) Energy loss function.	
	The x-dir and z-dir indicate the directions of the electric field polarized perpendicular and parallel to c-axis of the SL.}
	\label{Figure7}
\end{figure*}

\subsection{Electronic properties of the superlattice}

\begin{figure*}[ht]
	\centering
	\includegraphics[scale=0.6, angle=0] {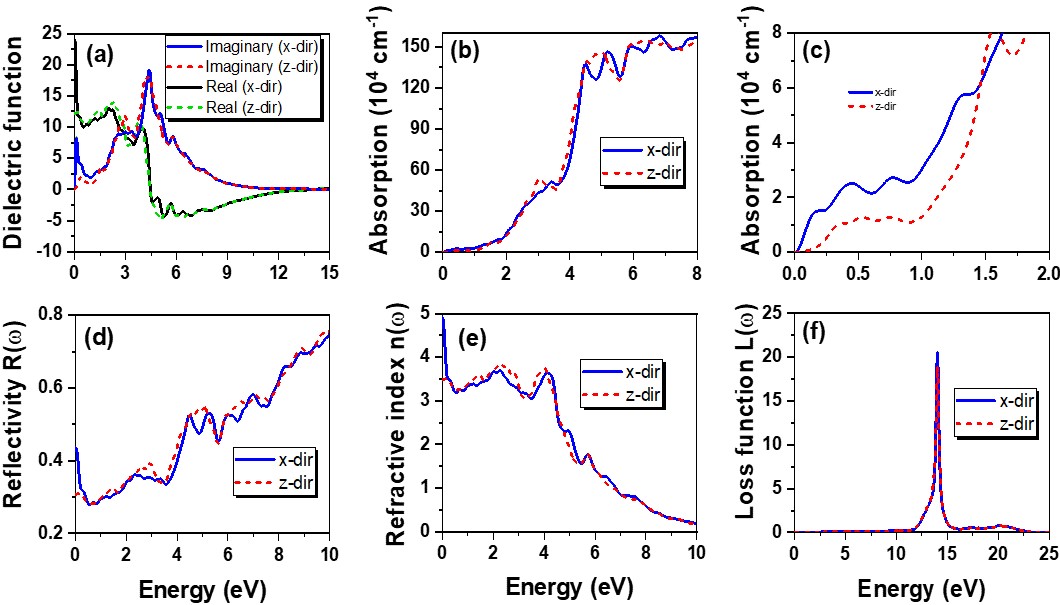}
	\caption{Frequency-dependent optical properties of InAs/InAs$_{0.625}$Sb$_{0.375}$ superlattice with lattice constant at T=300 K that is $a_{GaSb}$=6.0959 {\AA}. (a) Real and imaginary part of the dielectric function. (b)  Absorption spectra in the frequency-range between 0 and 8 eV. (c) Absorption spectra in the IR region. (d) Reflectivity. (e) Refractive index. (f) Energy loss function. The x-dir and z-dir indicate the directions of the electric field polarized perpendicular and parallel to c-axis of the SL.}
	\label{Figure8}
\end{figure*}

After the structural relaxation of the InAs/InAs$_{0.625}$Sb$_{0.375}$ SL, the electronic properties were calculated. Fig.~\ref{Figure6} shows the band structures calculated along the $\mathbf{k}$-path X-$\Gamma$-M-X for the InAs/InAs$_{0.625}$Sb$_{0.375}$ SL for the three lattice constants of bulk InAs, GaSb and AlSb, sequentially. We noticed a considerable decrease in the energy gaps compared to the parent compounds that allows their applicability in far-infrared detection. The value of band gap is calculated to be 116\, meV in the case of $a_{InAs}$, while for $a_{GaSb}$ this decreases to 87\, meV and further decreases to 53\, meV for $a_{AlSb}$, as shown in Table \ref{Table 3}. We noticed that, as the value of the lattice constant increases, the energy gap of the SL decreases.

At the $\Gamma$ point below the Fermi level, we have the highest valence band (HVB) two times degenerate. At 0.1 eV below HVB, we find another hole band two times degenerate, and another hole band two times degenerate at 0.25 eV below the HVB. All these six bands become almost degenerate at the high-symmetry point X, therefore, all these bands originate from the heavy-holes and light-holes of the bulk. We define the two highest valence bands as heavy-holes and the third highest valence band as light-hole. The presence of multiple heavy-hole bands is a huge difference respect to the bulk. 
The number of heavy-holes bands strongly depends on the period of the superlattice\cite{Manyk19vigo,RevModPhys.62.173}.
In the literature with tight-binding models, usually it was reported one heavy-hole band\cite{Manyk19vigo,Smith1987Proposal} for a 10.57 nm period. However, within first-principle calculations, we obtained two heavy-hole bands.  
The far we go from the Fermi level, heavier the effective masses of these bands become as opposite to the bulk where we have the HVB as heavy-hole and the second HVB as light-hole.
In the case of the superlattice, we define the HVB as heavy-hole1 and the second HVB as heavy-hole2.
To study the effect of lattice constant on the effective masses of the charge carriers, we computed the effective masses at the $\Gamma$ point of the SL such as  m$_{heavy-hole2}$, m$_{heavy-hole1}$ and m$_{electron}$ along the two in-plane directions i.e [100] and [110] as shown in Table \ref{Table 3}. 
We report different effective masses for the [100] and [110] directions, going beyond the approximation of isotropic effective mass in the k$_x$-k$_y$ plane found in the literature.

Due to the change of the symmetry, it is not possible to compare directly the effective masses along the same directions in bulk and SL. However, it is interesting to compare the effective masses of the SL respect to the effective masses of the bulk, since we expect that this trend can be observable in experiments. The biggest change in the SL is the reduction of  the effective masses of the HVB (heavy-hole1) respect to the bulk, while the second HVB (heavy-hole2) is now heavier as opposite to the bulk. The effective masses that we have calculated are in line with previous effective masses calculated with other theories\cite{Smith1987Proposal,Manyk19vigo,Tsai2020}. 
A considerable decrease of the effective masses along the [100] direction can be seen, as the lattice constant increases from 6.05830 {\AA} (InAs) to 6.1355 {\AA} (AlSb). Except for the m$_{electron}$, the effective masses along the [110] direction are  smaller than the effective masses along the [100] direction. 
We have noticed that the lattice constant not only influences the band gaps but also the effective masses associated with charge carries. Indeed, except for the m$_{electron}$, the effective masses decrease with the increasing of the lattice constant of the superlattice.\\

The SL Billouin zone is very different from the bulk Brillouin zone due to the change of the unit cell symmetries and lattice constants, in particular, the z-axis is extremely elongated flattening the band and increasing the effective mass along the [001] direction\cite{Manyk19vigo}. The calculated effective masses 
along [001] are reported in Table \ref{Table 4} and compared with the literature.
The calculated effective masses of the heavy-holes along [001] are extremely high as 54.9 which is comparable with the literature. The effective electron mass along [001] is 0.256 that is one order of magnitude larger than the in-plane effective masses. The reason of the difference between holes and electrons lies in the properties of the $\Gamma_6$ in zinc-blende semiconductors. The wave function of the $\Gamma_6$ state is in a quantum bonding state between the s-orbitals of Indium that can be approximated as:
$|\Gamma_6\rangle\approx\frac{1}{\sqrt{N}}\sum_{i=1,N} |{5s\text{-}In}_i\rangle$
where the index i runs on the N atoms of Indium present in the supercell, minor contributions from the s-orbitals of the anions are also present. This bonding state is delocalized in both side of the supercell.
The band bending of the T2SL InAs/InAs$_{0.625}$Sb$_{0.375}$ is shown in Fig. \ref{Figure6bis} together with the atomic character of the electrons of the lowest conduction band (LCB) and HVB of the InAs/InAs$_{0.625}$Sb$_{0.375}$ supercell. The DFT results confirm that the orbital character of the LCB is mainly 5s\text{-}In while the orbital character of the HVB is mainly 5p\text{-}Sb.
In summary, the electrons are delocalized along the superlattice owning a small mass, while the holes are localized in the In(As,Sb) side of the superlattice owning a huge effective mass. As a consequence the concentration and the distribution of the Sb doping mainly affects the hole carriers. The gap from the band bending is 0.13 eV are in agreement with the literature\cite{KROEMER2004196} and in qualitative agreement with the more accurate results for the supercell presented in Table \ref{Table 3}.\\

\begin{figure*}[ht]
	\centering
	\includegraphics[scale=0.6, angle=0] {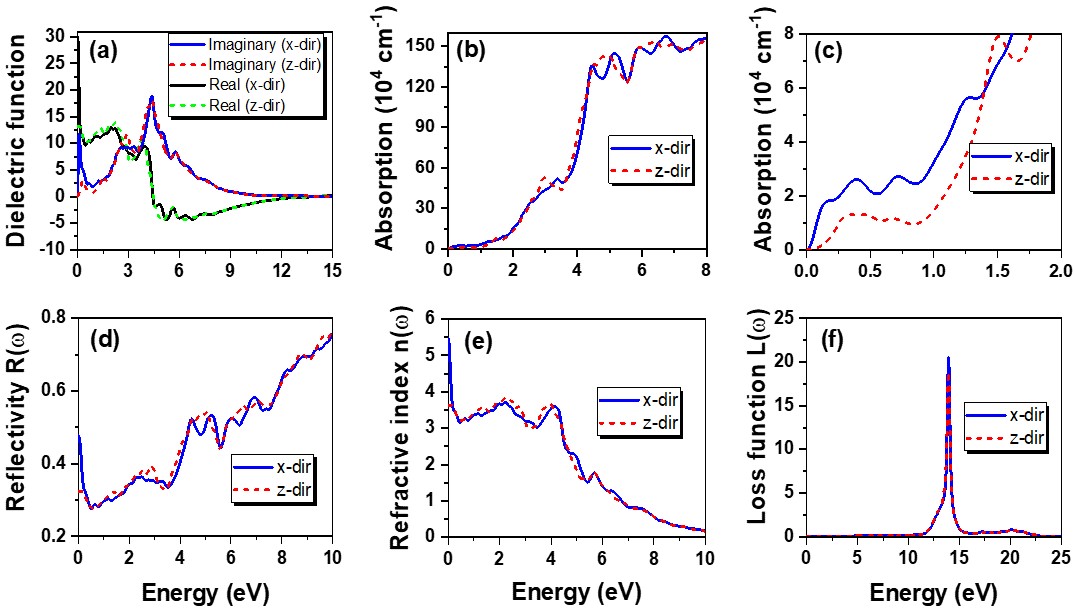}
	\caption{Frequency-dependent optical properties of InAs/InAs$_{0.625}$Sb$_{0.375}$ superlattice with lattice constant at T=300 K that is $a_{AlSb}$=6.1355 {\AA}. Real and imaginary part of the dielectric function. (b)  Absorption spectra in the frequency-range between 0 and 8 eV. (c) Absorption spectra in the IR region. (d) Reflectivity. (e) Refractive index. (f)  Energy loss function. The x-dir and z-dir indicate the directions of the electric field polarized perpendicular and parallel to c-axis of the SL.}
	\label{Figure9}
\end{figure*}

In Fig.~\ref{Figure6}(d), the interband electronic transitions related to the excitations in the SL are described. The light-matter interactions can be explained by the materials complex dielectric function, i.e  the imaginary part is directly linked to the absorption. The observed peaks in the inset of Fig.~\ref{Figure6}(d) can be connected to the excitation of electrons from valence band maximum to the conduction bands for $a_{InAs}$ based SL, which are also indicated by the interband transitions in the Fig. \ref{Figure6}(d) with the transition number 1 at 115\, meV and with the transition number 2 at 365\, meV. These excitons take place owing to the vertical transition from the valence band to the conduction bands at the $\Gamma$ point of the Brillouin zone as indicated in Fig.~\ref{Figure6}(d).   
Since the gap depends on the lattice constant, we figured out that also the energies required to cause the interband transitions are affected by the different lattice constants.

\subsection{Optical properties of the superlattice}

Based on the relaxed structure, we have explored the optical properties of InAs/InAs$_{0.625}$Sb$_{0.375}$ superlattices with the MBJ exchange-correlation functional. Again, the three different lattice constants are considered to calculate the optical properties as illustrated in Figs.~\ref{Figure7}, \ref{Figure8} and \ref{Figure9}, respectively. For the superlattice, we observed that all the optical properties are anisotropic because of the reduction in symmetry \cite{jiang2020effects}. Fig.s~\ref{Figure7}(a), \ref{Figure8}(a) and \ref{Figure9}(a) demonstrate the real and imaginary parts of the dielectric function for the three cases, respectively. Owing to the anisotropy, the values of static dielectric constants for SLs with the InAs, GaSb and AlSb lattice constant are (19.40, 15.57), (24.06, 12.20) and (30.00, 13.14) in the $x$ and $z$-directions, respectively. Table \ref{Table 5} demonstrates the static dielectric constants $(\varepsilon_{x}, \varepsilon_{z})$ and refractive indices (n$_{x}$, n$_{z}$) of InAs/InAs$_{0.625}$Sb$_{0.375}$ for lattice constants of InAs, GaSb and AlSb, respectively. The absorption coefficients of the SL for the three lattice constants are shown in Fig.s~\ref{Figure7}(b)(c), \ref{Figure8}(b)(c) and \ref{Figure9}(b)(c). The first absorption peak appears near ~0.115 eV and the second appears at ~0.365 eV in the case of InAs lattice constant, for the 
GaSb case they appear at ~0.092 eV and ~0.316 eV while for the AlSb case the first and second peaks appear at ~0.082 eV and ~0.273 eV, respectively. 

\begin{table}[th]
\centering
\caption{Static dielectric constants $(\varepsilon_{x}, \varepsilon_{z})$ and refractive indices (n$_{x}$, n$_{z}$) of InAs/InAs$_{0.625}$Sb$_{0.375}$ for lattice constants of InAs, GaSb and AlSb, respectively, using the MBJ approach.}
\setlength{\tabcolsep}{9 pt}
\renewcommand{\arraystretch}{1.5}
\scalebox{1.0}{
\label{Table 5}
\begin{tabular}{ccccc}
\hline
Lattice constant ({\AA}) & $\varepsilon_{x}$ & $\varepsilon_{z}$ & n$_{x}$ & n$_{z}$ \\
\hline
InAs (6.0583) & 19.4 & 15.6 & 4.40 & 3.40 \\
GaSb (6.0959) & 24.1 & 12.2 & 4.90 & 3.51 \\
AlSb (6.1355) & 30.0 & 13.1  & 5.51 & 3.60 \\
\hline
\end{tabular}}
\end{table}

In Figs. ~\ref{Figure7}(b), \ref{Figure8}(b) and \ref{Figure9}(b), we report the absorption spectra in the region between 0 and 8\,eV. The absorption spectra strongly increase from the frequency of the energy gap until approximately 5\,eV. Beyond 5\,eV, the absorption spectrum reaches a plateau with moderate oscillations of the order of 15\%. Including the electron-hole interaction, the plateau would be reached at lower frequencies.
In Figs. ~\ref{Figure7}(c), \ref{Figure8}(c) and \ref{Figure9}(c), we report the absorption spectra in the region between 0 and 2\,eV that is the relevant region for infrared detectors. We report the absorption coefficient when the electric field is polarized along the x-axis ($\alpha_{E_X}$) and z-axis ($\alpha_{E_Z}$). We can immediately see how the absorption coefficient is much larger when the electric field is polarized along the x-axis, i.e. orthogonal to the growth axis. Defining $\theta$ as the azimuthal angle between the polarization of the electric field and the z-axis, we have that the absorption coefficient $\alpha(\theta)$ is equal to:
\begin{equation}
\alpha(\theta)=\sin(\theta)\alpha_{E_X} + \cos(\theta)\alpha_{E_Z}    
\end{equation}

We have found characteristic absorption peaks in the low energy regions that quantify the characteristic interband transitions. This employs the possibility of using these SLs for infra-red detection.
Similar to the bulk case, the SL is also less reflective in the IR region (See Figs. ~\ref{Figure7}(d), \ref{Figure8}(d) and \ref{Figure9}(d)). Also, the refractive indices are higher in the IR range as shown in Figs~\ref{Figure7}(e), \ref{Figure8}(e) and \ref{Figure9}(e). Such less  reflection of the photons and  higher corresponding indices of refraction in the IR range of the electromagnetic spectrum suggest their applications in the IR detectors. It can be noticed that the static refractive indices $n_0$ are also anisotropic; these are (4.4, 3.4), (4.9, 3.5) and (5.5, 3.6) for InAs, GaSb and AlSb based lattice constants in the ($x$, $z$) directions (Table \ref{Table 5}). Moreover, the ELF for the three cases reveal almost identical behavior and show no energy loss up to 12\,eV, particularly for the photons in the IR range as shown in Figs. ~\ref{Figure7}(f), \ref{Figure8}(f) and \ref{Figure9}(f).\\

\begin{figure}[t]
	\centering
	\includegraphics[scale=0.64, angle=0] {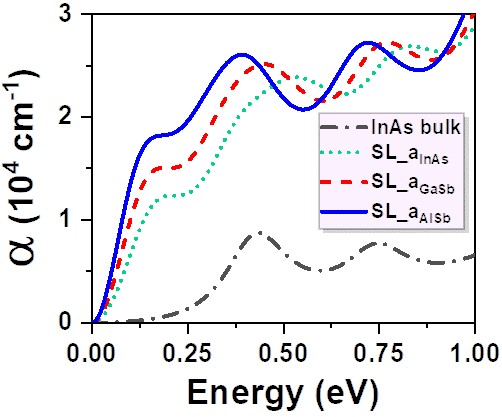}
	\caption{Absorption coefficient for an electric field polarized along the x-direction in the far-infrared regime between 0 and 1 eV. Spectra of InAs bulk (dash-dotted grey line) and InAs/InAs$_{0.625}$Sb$_{0.375}$ SLs for InAs (dotted green line), GaSb (dashed red line) and AlSb (solid blue line) substrate.}
	\label{Figure10}
\end{figure}

Since the most relevant physical property for the far-infrared detectors is the absorption spectrum, 
finally, we compare the absorption spectra in the IR region for InAs bulk and the SLs in Fig.~\ref{Figure10}. It can be seen that the absorption strongly increases for SL in the IR regime as compared to InAs bulk. Also, the absorption is observed to increase as a function of lattice constant and becomes maximum for the lattice constant of $a_{AlSb}$.
The reason of the large absorption coefficient of the SL could be attributed to the presence of the two heavy-hole bands. Then, we can speculate that thicker superlattice could produce even larger absorption spectra provided that the quality of the sample and the strain would be uniform along the superlattice. However, the uniform strain is limited by the critical thickness in the experimental realization.

To go beyond the independent-particle approximation employed here for the optical properties, we would need the using Bethe-Salpeter equation  that includes the electron-hole interaction.
The Bethe-Salpeter equation would produce a shift of the peak in the imaginary part to lower frequencies\cite{Sajjad18,VASP21Optical}.
Therefore, we expect that the lack of the electron-hole interaction would be responsible for a disagreement between theory and experiments at low frequencies. The Bethe-Salpeter equation would improve the agreement with experimental results,
however, the computational cost is not affordable for the large unit cell of the superlattice that we are describing.

\section{Conclusions}

We have investigated the electronic and optical properties of InAs and InSb bulk and their superlattices within MBJ approximation.
We focus on the InAs/InAs$_{0.625}$Sb$_{0.375}$ SL with the three lattice constants of the bulk InAs, GaSb and AlSb, respectively. It is observed that the electronic and optical properties effectively change due to the different symmetry present in the superlattice
and with the lattice constant of the superlattice. In the superlattice, we notice a considerable decrease in both the energy gaps and the effective masses of the heavy-holes compared to the bulk phases of the parent compounds. As opposite as the tight-biding literature, we have found two heavy-hole bands with the in-plane effective mass increasing as far as we go far from the Fermi level. 
In the k$_x$-k$_y$ plane, the effective masses of the heavy-holes in the superlattice becomes comparable with the one of the light-holes of the bulk. Along the growth axis, the effective mass of the hole becomes huge while the effective mass of electron increases less significantly. This happens since the electrons are delocalized in the entire superlattice while the holes are localized in the In(As,Sb) side of the superlattice.

Our theoretical calculations demonstrate that the absorption spectra in the far-infrared regime strongly increased in the case of SL respect to bulk InAs and InSb. The absorption coefficient of the SL is larger if the electric field is polarized in the direction orthogonal to the growth axis. Moreover, the absorption spectrum as a function of the lattice constant increased as a function of the lattice constant at low frequency. 
The appearance of multiple heavy-holes and a small energy gaps of the order of meVs produces an high absorption coefficient making these SLs employable for applications in far IR detectors.
The large sensitivity of the optical properties to structural and chemical degrees of freedom opens the possibility to engineering the optical property of the InAs-based superlattice making them even more appealing for the construction of far IR detectors.\\

\medskip

\section*{Acknowledgments}

G. H. and G. C. contributed equally to this work.
We thank T. Story and M. Birowska for useful discussions.
The work is supported by the Foundation for Polish Science through the International Research
Agendas program co-financed by the European Union within the Smart Growth Operational Programme.
This work was financially supported by the National Science Center
in the framework of the "PRELUDIUM" (Decision No.: DEC-2020/37/N/ST3/02338).
We acknowledge the access to the computing facilities of the
Interdisciplinary Center of Modeling at the University of
Warsaw, Grant No.~G75-10, G84-0, GB84-1 and GB84-7.
We acknowledge the CINECA award under the
ISCRA initiatives IsC81 "DISTANCE"
Grant, for the availability of high-performance computing
resources and support.

\bibliography{InAsInSb.bib}

\bibliographystyle{ieeetr}

\end{document}